\documentclass[10pt]{article}
\usepackage{amssymb,amsmath}
\def\RR{\mathbb{R}}

\def\CC{\mathbb{C}}

\def\ZZ{\mathbb{Z}}

\catcode`\@=11

\def\vereq#1#2{\lower3pt\vbox{\baselineskip1.5pt \lineskip1.5pt
\ialign{$\m@th#1\hfill##\hfil$\crcr#2\crcr\sim\crcr}}}

\def\vspace@{\def\vspace##1{\crcr\noalign{\vskip##1\relax}}}
\def\multilimits@{\bgroup\vspace@\Let@
 \baselineskip\fontdimen10 \scriptfont\tw@
 \advance\baselineskip\fontdimen12 \scriptfont\tw@
 \lineskip\thr@@\fontdimen8 \scriptfont\thr@@
 \lineskiplimit\lineskip
 \vbox\bgroup\ialign\bgroup\hfil$\m@th\scriptstyle{##}$\hfil\crcr}
\def\Sb{_\multilimits@}
\def\endSb{\crcr\egroup\egroup\egroup}
\def\Sp{^\multilimits@}

\newcommand{\be}[1]{\begin{equation}\label{#1}}
\newcommand{\ee}{\end{equation}}
\newcommand{\ba}[1]{\begin{eqnarray}\label{#1}}
\newcommand{\ea}{\end{eqnarray}}
\newcommand{\rf}[1]{(\ref{#1})}
\newcommand{\nn}{\nonumber}
\newcommand{\const}{\mbox{\rm const}}
\renewcommand{\Im}{\text{Im }}

\renewcommand{\#}{\sharp}

\newcommand{\diag}{\mbox{\rm diag}}

\newcommand{\vect}[1]{\mathbf{#1}}
\newcommand{\bsym}[1]{\boldsymbol{#1}}
\begin{document}
\title{Isospectrality of spherical  MHD dynamo operators:\\ Pseudo-Hermiticity and a no-go theorem}
\author{Uwe G\"unther\footnote{e-mail:
u.guenther@fz-rossendorf.de} \  and
 Frank Stefani\footnote{e-mail:  f.stefani@fz-rossendorf.de}\\
 Research Center Rossendorf, P.O. Box 510119, D-01314 Dresden, Germany}
 \date{7 August 2002}
\maketitle
\begin{abstract}
The isospectrality problem is studied for the operator
 of the spherical
hydromagnetic $\alpha^2-$dynamo. It  is shown that this operator
is formally pseudo-Hermitian ($J-$symmetric) and lives in a Krein
space. Based on the $J-$symmetry, an operator intertwining Ansatz
with first-order differential intertwining operators is tested for
its compatibility with the structure of the $\alpha^2-$dynamo
operator matrix. An intrinsic structural inconsistency is obtained
in the set of associated matrix Riccati equations. This
inconsistency is interpreted as a no-go theorem which forbids the
construction of isospectral $\alpha^2-$dynamo operator classes
with the help of first-order differential intertwining operators.
\end{abstract}
\section{Introduction}
The magnetic fields of stars and planets are generated by the
homogeneous dynamo effect in moving electrically conducting fluids
\cite{krause1}. This effect is explained within the framework of
magneto-hydrodynamics (MHD), but its experimental demonstration
was missing until recently. In 1999, the first successful dynamo
experiments in Riga and Karlsruhe \cite{dyn-1} opened up a new way for the
laboratory investigation of homogeneous dynamos. In connection
with the data analysis for the existing experiments and the design
of new dynamo experiments there is a growing interest in the
spectral properties of dynamos. Of particular interest is the
question whether isospectral dynamos can exist. First numerical
results on this topic were published in \cite{stefani2,stefani3}, but rigorous
results are still missing.

 As a step towards
clarification of this issue, we study in the present paper the
question  whether operator intertwining techniques from
 quantum mechanics (QM) can be adopted to MHD
dynamo models. In case of an affirmative answer we would
obtain an efficient tool for constructing isospectral classes of
MHD dynamo operators. Otherwise we would get a no-go theorem
which would forbid a straight analogy with quantum mechanical models.

Let us start by recalling some essentials of operator intertwining
transformations in QM \cite{raab}. Two operators $H_0$ and $H_1$
are said to be intertwined if there exist
 operators $A_+$ and $A_-$ so that
\begin{equation}\label{i1}
 H_1 A_+=A_+H_0\ , \quad  A_-H_1=H_0A_-  .
\end{equation}
For the corresponding eigenfunctions $\phi_0$ and $\phi_1$ holds,
up to normalization,
$$ \phi_1=A_+\phi_0\ , \quad A_-\phi_1=\phi_0 $$
and the operators $H_0$ and $H_1$ are isospectral, except for
those states that are annihilated by $A_+$ or $A_-$. In the case
of one-dimensional Schr\"odinger operators $H_0=p^2+V_0(x)$ and
$H_1=p^2+V_1(x)$ with the momentum operator given as
$p=-i\partial_x$, the intertwining operators can be chosen as
first order differential operators
\begin{equation}\label{i1a1} A_+=A:=ip+f, \quad
A_-=A^\dagger =-ip+f.
\end{equation}
 Structural compatibility of $H_0$ and $H_1$ with
the intertwining relations \rf{i1} requires that the function
$f(x)$ and the potentials $V_0(x)$, $V_1(x)$  are connected by
the consistency conditions
\begin{eqnarray}
V_1&=&V_0+2f^{\prime}\nn\\
-f^{\prime}+f^2&=&V_0-E\label{i1b.2}\\
f^{\prime}+f^2&=&V_1-E\label{i1b.3},
\end{eqnarray}
where the prime denotes
differentiation with respect to $x$; and $E$ is a constant of
integration. Linearization of the Riccati differential equations
\rf{i1b.2}, \rf{i1b.3} shows that this constant can be interpreted
as eigenvalue of the Schr\"odinger operators $H_0$ and $H_1$
\begin{eqnarray}
H_0\chi_0=E\chi_0 &\qquad \mbox{\rm for}& f=-\frac{\chi_0 ^{\prime}}{\chi_0},\label{i1c.1}\\
H_1\chi_1=E\chi_1 &\qquad \mbox{\rm for}& f=\frac{\chi_1 ^{\prime}}{\chi_1},\label{i1c.2}
\end{eqnarray}
where $\chi_0$ and $\chi_1$ are formal, and not necessarily
normalized eigenfunctions of $H_0$ and $H_1$, respectively. They
are connected by the product relation
\begin{equation}\label{i1cc}
\chi_0\chi_1=c
\end{equation}
with $c$ a  non-vanishing constant.
 It is straightforward to verify that the shifted
Schr\"odinger operators are factorizable in terms of the
intertwining operators $$ H_0-E=A^\dagger A, \quad H_1-E=AA^\dagger . $$
First-order differential intertwining transformations of type
\rf{i1a1} are known as Darboux transformations \cite{matveev} and
are widely used to generate isospectral operator classes from
given operators with known spectra \cite{raab,darb2,iso-riemann}.
In particular, intertwining constructions are a basic ingredient
of super-symmetric quantum mechanical models \cite{raab,susy1} and
their generalizations to pseudo-supersymmetric systems
\cite{most1,most4}. As it was demonstrated in \cite{geszt1}, a
double-intertwining (double commutation) method can provide a tool
for inserting additional eigenvalues in spectral gaps of given
background Schr\"odinger and Jacobi operators.

Motivated by the large number of exact results on isospectral
 classes obtained by operator intertwining constructions, it is
 natural
 to investigate whether  MHD dynamo operators are also
  suitable for this technique. For this purpose we study in the
  present paper the simplest mean-field MHD dynamo configuration --- the
spherical $\alpha^2-$dynamo
 \cite{krause1}. In terms of the radial momentum operator $p=-i(\partial_r +1/r)$
 the  $2\times
 2$ operator matrix of the $\alpha^2-$dynamo is given
 as\footnote{A brief outline of the derivation of the $\alpha^2-$dynamo operator matrix from the MHD mean-field
 induction
 equation can be found in Appendix \ref{induct1}.}
\begin{equation}\label{i2} \hat H_l[\alpha]  \equiv \left(
\begin{array}{rrr}
  {-p^2 -\frac{l(l+1)}{r^2}} &  & {\alpha (r)} \\ \\
    {p \alpha (r)p   +\alpha (r)\frac{l(l+1)}{r^2}} &  & {-p^2  -\frac{l(l+1)}{r^2}} \\
 \end{array}\right)
\end{equation}
and lives on the domain \begin{align*}
 {\cal D}(\hat H_l[\alpha])&:=\left\{
\psi = \left(\begin{array}{r}\psi_1 \\ \psi_2 \end{array}\right):
\ \psi\in \tilde{{\cal H}}\equiv {\cal H} \oplus {\cal H}, \
{\cal H}= L_2(\Omega,r^2 dr),\right. \\
& \left. \Omega =[0,1], \ \psi (1)=0, \
\left.r\psi(r)\right|_{r\to 0}\to 0
 \right\}
\end{align*}
 in the Hilbert space $\tilde{{\cal H}}$.
It describes the coupled $l-$modes of the poloidal and toroidal
magnetic field components in a mean-field dynamo model with
helical turbulence function $\alpha (r)$. The function $\alpha(r)$
does not depend on $l$ and
we assume that it is real-valued, positive definite, bounded,
and sufficiently smooth on $\Omega$: $\Im \alpha =0, \  0<
\alpha \le c_1<\infty, \ \alpha \in C^4(\Omega)$.
 The idealized
 boundary condition $\psi (r=1)=0$  corresponds to a super-conducting
spherical boundary shell and is chosen here to ensure simplicity
of the subsequent considerations \cite{proctor1}. For more
realistic models with close relation to stellar dynamos, the
spherical fluid configurations confined to $r<1$ can be assumed as
embedded in empty space. The
 boundary condition should then be replaced by  $\left.\hat B_l\psi\right|_{r=1}
=0$ with $\hat B_l=\diag [\partial_r+(l+1)/r,1]$ \ (see e.g. \cite{krause1}) what will require a more general approach
than that presented in the present paper.

Exploring the fundamental symmetry of the $\alpha^2-$dynamo
operator matrix  we find in section \ref{hermiticity} that $\hat
H_l[\alpha]$ acts as a symmetric operator on the Hilbert space
$\tilde{{\cal H}}$ when this is endowed with an indefinite metric
$J$. I.e., the $\alpha^2-$dynamo operator matrix is a
$J-$symmetric (formally $J-$self-adjoint) operator
$$
 \hat
H_l[\alpha]=\hat H^{\#}_l[\alpha]:= J\hat H^\dagger_l[\alpha]J
$$
  living in a Krein
space $\tilde{{\cal K}}=\tilde{{\cal H}}_J$ \cite{azizov}.
$J-$self-adjointness is a natural property of operators from
different fields of physics. Examples are, e.g., the
super-symmetric Dirac operator \cite{LT}, $PT-$symmetric
non-Hermitian Hamiltonians in QM \cite{most1,bender} as well as
the Wheeler-DeWitt operator for a cosmological
Friedman-Robertson-Walker model coupled to a real massive scalar
field
 \cite{most1}. Since the recent
paper
 series \cite{most1,most4,most2} of A. Mostafazadeh on
non-Hermitian operators with real spectra,  $J-$self-adjoint operators  are also known as
pseudo-Hermitian operators.

In analogy with the simple quantum mechanical model described
above,  we base our isospectrality analysis on an intertwining
Ansatz for two $\alpha^2-$dynamo operators with helical
turbulence functions $\alpha_0(r)$, \ $\alpha_1(r)$
$$
\hat H_{l_0}[\alpha_0]-E I=-\hat A \hat A^{\#}, \qquad \hat
H_{l_1}[\alpha_1]-E I=-\hat A^{\#} \hat A
$$
and
intertwining operator matrices $\hat A$, $\hat A^{\#}$ that are
first-order
 differential operators
$$
 \hat A := i R(r)p+Q(r), \qquad \hat A^{\#}:=-i p
R^{\#}(r)+Q^{\#}(r)\; .
$$
This Ansatz leads to a set of six
consistency conditions on the matrices $R(r)$ and $Q(r)$ which are
studied in section
 \ref{riccati}. It is shown that one pair of  conditions fixes the structure of $R(r)$
 in terms of the helical turbulence functions $\alpha_0(r)$ and $\alpha_1(r)$. A second pair
is equivalent to the symmetry relations $B=B^{\#}$, $U=U^{\#}$ on
the matrix functions
\begin{eqnarray*}
 B&:=&R^{\#}Q \\
 U&:=&R[Q^{\#}-(R^{\#})^{\prime}]=RBR^{-1}-R(R^{\#})^{\prime}
 \end{eqnarray*}
and can be regarded as  an implicit consequence of the
$J-$pseudo-Hermiticity of the operator matrices $\hat
H_{l_0}[\alpha_0]$ and $\hat H_{l_1}[\alpha_1]$. (The prime denotes the derivative with respect to $r$.) The remaining two
conditions can be transformed into a pair of coupled matrix
Riccati
 differential equations (MREs) on $B$ and $U$.

The consistency of the six conditions is analyzed in section
\ref{no-go} with the help of a step-by-step reduction of their
complexity. First, we conclude from the limiting behavior of the
MREs for $r\to 0$
  that the angular mode numbers $l_0$ and $l_1$ in the two dynamo operator
 matrices should be connected by the incremental relation
$l_1=l_0+1$.  Then we use the $J-$symmetry of $B$ to derive from
the coupled MREs a system of coupled non-linear ordinary
differential equations (ODEs) involving the helical turbulence
functions $\alpha_0(r)$ and $\alpha_1(r)$. Analyzing these ODEs we
are able to show the existence of an inherent contradiction
between them. As implication, we arrive at a no-go theorem which
states that the six consistency conditions cannot be fulfilled
simultaneously and that, hence, the structure of the
$\alpha^2-$dynamo operator matrices is not suitable for an
operator intertwining technique based on an Ansatz with
first-order differential intertwining operators.

In the concluding section \ref{conclu} we briefly discuss some
other methods which could be useful for studying isospectrality
issues of the dynamo operator matrix and which possibly could
provide a technique to construct classes of isospectral spherical
$\alpha^2-$dynamo operators.

\section{$J-$Symmetry of the dynamo operator matrix\label{hermiticity}}
In this section, we study the fundamental
symmetry $J$ of the $\alpha^2-$dynamo operator matrix \rf{i2} what
allows us to choose an appropriate  Ansatz for the intertwining
operators $A_+$ and $A_-$.

We start our consideration by introducing the auxiliary operator
$$
Q[\alpha]:=p\alpha p + \alpha \frac{l(l+1)}{r^2}
$$
defined on the domain
$$
{\cal D}(Q)= \left\{ \phi: \ \phi\in {\cal H} = L_2(\Omega,r^2 dr), \
\phi (1)=0, \ \ \left.r\phi(r)\right|_{r\to 0}\to 0
 \right\} \nn
$$
in the Hilbert space ${\cal H}$. The operator $Q[\alpha]$ is a
formally self-adjoint\footnote{In the subsequent compatibility
analysis of the operator intertwining construction we restrict our
attention to symmetric (formally self-adjoint) operators. For
simplicity, we leave questions of self-adjoint extensions and
corresponding generalized boundary conditions \cite{albev,gorb}
for the bi-component functions $\psi$ aside.} singular
differential operator $Q=Q^\dagger$ which acts as symmetric operator on
${\cal H}$. In terms of $Q[\alpha]$ the dynamo operator matrix and
its formal adjoint read
$$
\hat H_l[\alpha]=\left(\begin{array}{cc}-Q[1] & \alpha\\ Q[\alpha] & -Q[1]
 \end{array}\right)\ , \qquad
\hat H_l^\dagger[\alpha]=\left(\begin{array}{cc}-Q[1] & Q[\alpha]\\ \alpha & -Q[1]
 \end{array}\right) \nn
$$
so that the fundamental (canonical) symmetry can be obtained as
\begin{equation}\label{4}
\hat H_l[\alpha]=\hat H_l^{\#}[\alpha]:=J\hat H_l^\dagger[\alpha]J , \qquad J=\left(\begin{array}{cc}0 & 1\\ 1 & 0
 \end{array}\right).
\end{equation}
Diagonalizing the matrix $J$
$$
J\longrightarrow \eta =S^TJS, \quad
S=\frac{1}{\sqrt{2}}\left(\begin{array}{rr}1&-1\\1&1\end{array}\right),
\quad \eta =\left(\begin{array}{rr}1&0\\0&-1\end{array}\right)
$$
we see that $\hat H_l[\alpha]$
is equivalent to the operator matrix
$$
\check{H}_l[\alpha] =S^T\hat H_l[\alpha]S=\frac12\left(\begin{array}{ccc}Q[\alpha
-2]+\alpha & & -Q[\alpha] + \alpha
 \\ Q[\alpha] - \alpha &&Q[-\alpha -2]-\alpha\end{array}\right)
$$
with the property $\check{H}_l[\alpha]=\eta \check{H}_l[\alpha]^\dagger \eta$.
The fundamental $\eta -$symmetry of the operator matrix $\check{H}_l[\alpha]$
implies that ${\cal D}(\check{H}_l)$ could be endowed with
the indefinite  metric $\eta$ so that $\check{H}_l[\alpha]$ becomes a symmetric operator
on ${\cal D}(\check{H}_l)$.
Due to the invariance of the signature under the
transformation $S$ the domain  ${\cal D}(\hat H_l)$ can also be endowed with a natural indefinite
inner product $\left[ \cdot,\cdot \right]_J$ defined by the metric $J$
$$
\left[ x,y\right]_J := (x,J y),  \quad x,y\in
\tilde{{\cal H}}={\cal H} \oplus {\cal H},
$$
where $(\cdot,\cdot)$ denotes the usual inner (scalar) product in
the Hilbert space $\tilde{{\cal H}}$. This means that $\hat
H_l[\alpha]$ is a $J-$symmetric operator which acts as symmetric
operator in a Krein space\footnote{For surveys on operators in
Krein spaces (Hilbert spaces with additional indefinite inner
product structures) we refer to the mathematical literature
\cite{azizov,L2}.} $\tilde{{\cal K}}=\tilde{{\cal H}}_J$
$$
\left[\hat H_l x,y\right]_J=\left[ x,\hat
H_l^{\#} y\right]_J.
$$
From its operator-matrix representation \rf{4} we see that $J$ is self-adjoint, involutory and unitary
$$
J^\dagger=J, \quad J^2=I, \quad J^{-1}=J^\dagger
$$
so that $\hat H_l[\alpha]$ is a $J-$pseudo-Hermitian operator in the sense of
\cite{most1,most4,most2}.

The eigenvalues of $J-$pseudo-Hermitian operators are known \cite{most1,azizov,L2}
to be either real or to come in complex-conjugate pairs. Here we
illustrate this property by passing from the eigenvalue problem
for
the linear operator pencil
$$
\hat L_l[\alpha,\lambda]\psi :=\left(\hat
H_l[\alpha]-\lambda\right)\psi=0
$$
to the eigenvalue problem $L_l[\alpha,\lambda]\psi_1=0$ for the associated quadratic operator pencil $L_l[\alpha,\lambda]$.
This pencil can be derived explicitly from the Ansatz
$$
\psi = \left(\begin{array}{r}\psi_1\\ \frac1\alpha [Q(1)+\lambda ]\psi_1
\end{array}\right)
$$
with
$\alpha(r)\neq 0$.  As result we obtain
\begin{eqnarray*}
L_l[\alpha,\lambda]\psi_1 &\equiv
&\left\{\left[Q[1]+\lambda\right]\frac1\alpha
\left[Q[1]+\lambda\right]-Q[\alpha] \right\}\psi_1 =0 \nn \\ &=&
(A_2\lambda^2+A_1\lambda+A_0)\psi_1=0.
\end{eqnarray*}
The operators
$$
A_0:=Q[1]\frac1\alpha Q[1]-Q[\alpha] , \qquad
A_1:=Q[1]\frac1\alpha +\frac1\alpha Q[1] , \qquad
A_2:=\frac1\alpha
$$
are formally self-adjoint on ${\cal D}(Q)$ so that the
functionals $a_j:=(A_j\psi_1,\psi_1), \ j=1,2,3$ \
are real-valued: $\Im a_j=0$.

From the quadratic equation
$$
(L_l[\alpha,\lambda]\psi_1,\psi_1)=a_2\lambda^2+a_1\lambda +a_0=0
$$
 we conclude that the
eigenvalues of the $J-$pseudo-Hermitian dynamo operator matrix
$\hat H_l$ and its associated pencil $L_l$ occur as eigenvalue
pairs \cite{markus}
$$ \lambda_\pm=\frac{1}{2a_2}\left(
-a_1\pm\sqrt{a_1^2-4a_0a_2}\right). $$ Obviously, the
 sign of the discriminant $\Delta:=a_1^2-4a_0a_2$ defines whether $\lambda _\pm$ are both real or
 pairwise
 complex conjugate. The transition from real eigenvalues $\lambda_\pm$ to
 complex ones occurs at $\Delta=0$ where the eigenvalue becomes two-fold degenerate
  $\lambda_+=\lambda_-=\lambda_0=-\frac{a_1}{2a_2}$.
This general behavior of $\lambda_\pm$ confirms the results of numerical simulations
\cite{stefani3}, which showed that a scaling of the helical turbulence
function $\alpha$ leads to a pairwise intersection of real eigenvalue
branches of $\hat H_l$ and a transition at the intersection points to a pair of complex conjugate eigenvalues.

We note that  at the two-fold degenerate points $\lambda_0$ of the
spectrum with $\Delta=0$ a Jordan-Keldysh chain
\cite{kato} exists for the linear operator pencil
$$
\hat L_l(\lambda_0)\psi=0\ , \qquad \hat L_l(\lambda_0)\chi=\psi
$$
as well as for the quadratic operator pencil
$$
L_l(\lambda_0)\psi_1=0\ , \qquad L_l(\lambda_0)\chi_1 +
\partial_{\lambda}L_l(\lambda_0)\psi_1=0.
$$
Both are built up from eigenvectors $\psi$, $\psi_1$ and associated vectors $\chi$, $\chi_1$, respectively.

\section{Consistency conditions and matrix Riccati equations\label{riccati}}
 The fundamental $J-$symmetry
($J-$pseudo-Hermiticity) of the $\alpha^2-$dynamo operator matrix
provides a natural Ansatz for an intertwining construction which
respects this symmetry:
\begin{equation}\label{r1}
\hat H_{l_0}[\alpha_0]-E I=-\hat A \hat A^{\#}, \qquad
\hat H_{l_1}[\alpha_1]-E I=-\hat A^{\#} \hat A .
\end{equation}
In general, the operator matrix $\hat A$ could be an $n$th-order differential
operator of the form
$$
\hat A=\sum^n_{k=1}R_k(r)(i p)^k+Q(r)
$$
with $2\times 2$ matrices $R_k(r)$ and $Q(r)$ as coefficients. For
simplicity, we restrict our attention in the present paper to the
first-order differential operator
\begin{equation}\label{r3}
\hat A=i R(r)p+Q(r)
\end{equation}
with $J-$adjoint
$A^{\#}=-ipR^{\#}(r)+Q^{\#}(r)$.  Here we define the $\#-$operation
for a given $2\times 2$ matrix $C$ as
$$
C^{\#}=JC^\dagger J=J C^{\ast T}J.
$$
Asterisk and superscript "${}^T$"
denote complex conjugation and transposition, respectively.

Let us introduce the abbreviations
\begin{eqnarray*}
K_{0,1}&:=&I-\alpha_{0,1}\sigma_-\label{r7.1} \nn\\
M_{0,1}&:=&K_{0,1}\frac{l_{0,1}(l_{0,1}+1)}{r^2}+E I-\alpha_{0,1}\sigma_+\label{r7.2}\nn
\end{eqnarray*}
with the nilpotent matrices $\sigma_{\pm}$ defined as
$\sigma_+:=\left(\begin{array}{rr}0&1\\0&0\end{array}\right)$, \
$\sigma_-:=\left(\begin{array}{rr}0&0\\1&0\end{array}\right)$. The
shifted $\alpha^2-$dynamo operator matrices in \rf{r1} take then
the short form
\begin{equation}\label{r5}
\hat H_{l_{0,1}}[\alpha_{0,1}]-E I=-pK_{0,1}p-M_{0,1}.
\end{equation}
Substituting \rf{r3} and \rf{r5} into the intertwining Ansatz
\rf{r1}, making use of commutation relations like
$[p,R(r)]=-iR^{\prime}(r)$ and equating the coefficient matrices
of the $p^2$, $p$, $I$ terms we obtain the following six consistency
conditions
\begin{align}
\hat H_{l_0}:\qquad &p^2:&RR^{\#}&=K_0\label{r6.1}\\
&p:&RQ^{\#}-QR^{\#}-R(R^{\#})^{\prime}+R^{\prime}R^{\#}&=0\label{r6.2}\\
&I:&QQ^{\#}-R(R^{\#})^{\prime\prime}+R(Q^{\#})^{\prime}-Q(R^{\#})^{\prime}
&=M_0\label{r6.3}\\ \nonumber\\ \hat H_{l_1}:\qquad
&p^2:&R^{\#}R&=K_1\label{r6.4}\\
&p:&-R^{\#}Q+Q^{\#}R&=0\label{r6.5}\\
&I:&Q^{\#}Q-\left(R^{\#}Q\right)^{\prime}&=M_1.\label{r6.6}
\end{align}
For a successful intertwining construction these matrix equations
should be fulfilled simultaneously. So, the main task consists in
finding explicit solution sets for \rf{r6.1} - \rf{r6.6}.
Alternatively,  we should obtain intrinsic contradictions within
this equation system which could be interpreted as a no-go theorem
forbidding this construction for $\alpha^2-$dynamo operator
matrices.

We start our analysis with eqs. \rf{r6.1} and \rf{r6.4}. From the
tautologies \ $RR^{\#}R=RR^{\#}R$ \ and \
$R^{\#}RR^{\#}=R^{\#}RR^{\#}$ \ follows
$$
RK_1=K_0R,\qquad K_1R^{\#}=R^{\#}K_0
$$
what with
$$
R=\left(\begin{array}{rr}r_{11}&r_{12}\\r_{21}&r_{22}\end{array}\right),
\qquad R^{\#}=\left(\begin{array}{rr}r^\ast_{22}&r^\ast_{12}\\
r^\ast_{21}&r^\ast_{11}\end{array}\right)
$$
 yields
\begin{equation}\label{r10}
r_{12}=0,\qquad
\frac{\alpha_1}{\alpha_0}=\frac{r_{11}}{r_{22}}=\frac{r^\ast_{11}}{r^\ast_{22}}.
\end{equation}
Hence, we can set
$$
r_{11}=|r_{11}|e^{i\gamma}, \qquad r_{22}=|r_{22}|e^{i\gamma},
\qquad r_{21}=|r_{21}|e^{i(\gamma +\varepsilon)}.
$$
Using this and \rf{r10} in
$$
RR^{\#}=K_0=\left(\begin{array}{cc}1&0\\-\alpha_0&1\end{array}\right),
\qquad
R^{\#}R=K_1=\left(\begin{array}{cc}1&0\\-\alpha_1&1\end{array}\right)
$$
we find
\begin{equation}\label{r12}
R=e^{i\gamma}\left(\begin{array}{cc}\sqrt{\frac{\alpha_1}{\alpha_0}}&0\\-\frac
12 \sqrt{\alpha_0\alpha_1}\left(1+i\tan \varepsilon\right)
&\sqrt{\frac{\alpha_0}{\alpha_1}}\end{array}\right),
\end{equation}
where the phases $\gamma$ and $\varepsilon$ are still undefined.

As next step we analyze  Eqs. \rf{r6.2} and \rf{r6.5}. It is easily
seen that defining the matrices
\begin{equation}\label{r13}
U:=R\left[Q^{\#}-(R^{\#})^{\prime}\right] ,\qquad B:=R^{\#}Q
\end{equation}
these equations are equivalent to the $J-$symmetry relations
$$
U=U^{\#},\qquad B=B^{\#}.
$$
Due to the different symmetry content of $B$ and $Q$ it is natural
to consider $B$ as primary structural element of the intertwining
construction, and $Q$ as a secondary one. So, we perform our
subsequent investigation in terms of $B$ and $R$.
Explicitly, the $J-$symmetry is realized by the matrix structure
\begin{equation}\label{r15}
B=\left(\begin{array}{cc}b_1+ib_4&b_2\\b_3&b_1-ib_4\end{array}\right),
\qquad \Im b_k=0, \quad k=1,\ldots,4 \ .
\end{equation}
Furthermore, we  exclude $Q$ from \rf{r13} to obtain
\begin{equation}\label{r16}
U=RBR^{-1}-R(R^{\#})^{\prime}.
\end{equation}
Introducing the notation $N:=R^{-1}R^{\prime} $ and substituting
\rf{r16} into the symmetry relation $U=U^{\#}$ yields the
additional constraint
\begin{equation}\label{r18}
\left[B,K_1^{-1}\right]=N^{\#}-N.
\end{equation}
From eq. \rf{r12}  we find
\begin{eqnarray}
N&=& i\gamma^{\prime}I+\left(\begin{array}{cc}-q&0\\
f&q\end{array}\right)\label{r19.1} \nn \\
q&=&\frac 12
\left(\frac{\alpha_0^{\prime}}{\alpha_0}-\frac{\alpha_1^{\prime}}{\alpha_1}\right)\label{r19.2}\\
f&=&-\frac{\alpha_1}{2}\left[
\frac{\alpha_0^{\prime}}{\alpha_0}\left(1+i\tan \varepsilon
\right) +i\frac{\varepsilon^{\prime}}{\cos^2
\varepsilon}\right]\label{r19.3}
\end{eqnarray}
so that  \rf{r18} transforms to
$$
\alpha_1\left(\begin{array}{cc}b_2&0\\
-2ib_4&-b_2\end{array}\right)=-2i\gamma^{\prime}I+
\left(\begin{array}{cc}2q&0\\ f^\ast- f&-2q\end{array}\right).
$$
Finally, we arrive at the following restrictions on the phase
$\gamma $ and the components $b_2$ and $b_4$ of the matrix $B$:
\begin{equation}\label{r21}
\gamma^{\prime} =0, \qquad b_2=\frac{2q}{\alpha_1}, \qquad b_4=\frac{\Im f}{\alpha_1}=-\frac 12\left(
\frac{\alpha_0^{\prime}}{\alpha_0}\tan \varepsilon
+\frac{\varepsilon^{\prime}}{\cos^2 \varepsilon}\right).
\end{equation}

Summarizing the implications of the first four consistency
conditions we see that they are free of intrinsic contradictions.
From the initially eight arbitrary complex-valued functions
contained in the matrices $R$ and $Q$ only the three real-valued
functions $(b_1, b_3, \varepsilon)$ are still undefined. Together
with the helical turbulence functions $(\alpha_0, \alpha_1)$ and
the constants $(\gamma,E,l_0,l_1)\in \RR^2\times\ZZ_+^2$ we
expect them to be highly fine-tuned by the remaining two
consistency conditions \rf{r6.3} and \rf{r6.6}.

Let us study these conditions now. Making use of the definitions
of  $U$ and $B$ in \rf{r13}, their implications
\begin{eqnarray}
Q^{\#}-(R^{\#})^{\prime}&=&R^{-1}U,\label{r22.1}\\
(Q^{\#})^{\prime}-(R^{\#})^{\prime\prime}&=&-R^{-1}R^{\prime}R^{-1}U+R^{-1}U^{\prime},\label{r22.2}
\nn\\
Q&=&(R^{\#})^{-1}B \label{r22.3}
\end{eqnarray}
and setting at the end $RR^{\#}=K_0$, $R^{\#}R=K_1$ according to eqs.
\rf{r6.1}, \rf{r6.4}, we find that \rf{r6.3} and \rf{r6.6} transform
to the matrix Riccati equations (MREs)
\begin{eqnarray}
U^{\prime}&=&M_0-UK_0^{-1}U,\label{r23.1}\\
B^{\prime}&=&-M_1+BK_1^{-1}B\label{r23.2}.
\end{eqnarray}

Similar to the linearization of the scalar Riccati equations mentioned in \rf{i1b.2} - \rf{i1c.2} of the Introduction,
the MREs \rf{r23.1}, \rf{r23.2} can be linearized by an Ansatz
\cite{grass1,wint1}
\begin{eqnarray}
U&=&VW^{-1},\quad V,W\in \CC^{2\times 2}, \quad \det (W)\neq
0,\label{r24.1}\\ B&=&XY^{-1},\quad X,Y\in \CC^{2\times
2},\quad \det (Y)\neq 0\label{r24.2}.
\end{eqnarray}
As result
we arrive at the equation systems
\begin{equation}\label{r25}
\left(\begin{array}{c}V^{\prime}\\
W^{\prime}\end{array}\right)=\left(\begin{array}{cc}0&M_0\\
K_0^{-1}&0\end{array}\right)\left(\begin{array}{c}V\\
W\end{array}\right),\qquad
\left(\begin{array}{c}X^{\prime}\\
Y^{\prime}\end{array}\right)=-\left(\begin{array}{cc}0&M_1\\
K_1^{-1}&0\end{array}\right)\left(\begin{array}{c}X\\
Y\end{array}\right).
\end{equation}
The $4\times 2$ matrices $\left(\begin{array}{c}V\\
W\end{array}\right),\left(\begin{array}{c}X\\
Y\end{array}\right)\in \CC^{4\times 2}$ are defined up to  $GL(2,\CC)\times
GL(2,\CC)-$transformations
$$
\left(\begin{array}{c}\tilde V\\ \tilde
W\end{array}\right)=\left(\begin{array}{c}VG_0\\
WG_0\end{array}\right),\qquad \left(\begin{array}{c}\tilde X\\
\tilde Y\end{array}\right)=\left(\begin{array}{c}XG_1\\
YG_1\end{array}\right), \qquad G_0,G_1\in GL(2,\CC)
$$
and can be interpreted as homogeneous coordinates of two points on
a complex Grassmann manifold $G_2(\CC^{4})$ which consists of
 $2-$dimensional complex subspaces in $\CC^{4}$ (see, e.g.
\cite{grass1,wint1}). The matrices $U=VW^{-1}$ and $B=XY^{-1}$ are the
corresponding affine coordinates of these points.

Differentiating \rf{r25} and substituting $V=K_0W^{\prime}$,
$X=-K_1Y^{\prime}$ it is easily seen that the equation systems
\rf{r25} are equivalent to the second-order matrix differential
equations
\begin{eqnarray}
\left(\partial_rK_0\partial_r-M_0\right)W&=&0,\label{r27.1}\nn\\
\left(\partial_rK_1\partial_r-M_1\right)Y&=&0. \label{r27.2}
\end{eqnarray}
This implies that the matrices $\tilde W=r^{-1}W$, $\tilde Y=r^{-1}Y$
should be
formal (non-normalized)
solutions of the eigenvalue equations for the dynamo operator matrices $\hat H_{l_0}[\alpha_0]$, $\hat H_{l_1}[\alpha_1]$,
respectively
$$
\hat H_{l_0}[\alpha_0]\tilde W=E\tilde W,\qquad \hat
H_{l_1}[\alpha_1]\tilde Y=E\tilde Y.
$$
A comparison with the simple QM model from the Introduction
shows that the intertwining operator matrix $\hat A$ should be expressible in
terms of $W$ or $Y$, and that $W$ and $Y$ should be connected by a
product invariant like \rf{i1cc}. With the help of \rf{r22.1},
\rf{r22.3} and \rf{r25} we find
$$
\hat
A=R\left(ip-Y^{\prime}Y^{-1}\right)=\left(ip+K_0W^{\prime}W^{-1}K_0^{-1}\right)R.
$$
In order to obtain the product invariant which connects $W$ and
$Y$, we use a slightly modified version of \rf{r16}
$$
U=RB^{\#}R^{-1}-R(R^{\#})^{\prime}
$$
and substitute from \rf{r24.1} - \rf{r25}
$$
U=RR^{\#}W^{\prime}W^{-1}, \qquad
B^{\#}=-(Y^{\#})^{-1}(Y^{\#})^{\prime}R^{\#}R
$$
so that
\begin{equation}\label{r32}
W^{\prime}W^{-1}=-(R^{\#})^{-1}(Y^{\#})^{-1}(Y^{\#})^{\prime}R^{\#}-(R^{\#})^{-1}(R^{\#})^{\prime}.
\end{equation}
This equation is of the type $g=g_1 n$, \ $(\partial_r
g)g^{-1}=(\partial_r g_1)g^{-1}_1+g_1(\partial_r
n)n^{-1}g_1^{-1}$.  Hence, integration of \rf{r32} yields the
product invariant
$$
Y^{\#}R^{\#}W=C, \qquad \det(C)\neq 0,
$$
with $C$  a constant non-singular matrix.

So far, we have obtained a $1:1$ generalization of the
intertwining technique from the simple QM example described in the Introduction to our $J-$symmetric dynamo operator
model. It remains to test whether the MREs of this model are
consistent. This will be the subject of the next section.

\section{No-go theorem\label{no-go}}

In order to test the pair of MREs  \rf{r23.1}, \rf{r23.2} for
consistency, we make use of \rf{r16}, \rf{r18} as well as the relation
$$
N+K_1^{-1}N^{\#}K_1=K_1^{-1}K_1^{\prime}=K_1^{\prime}
$$
  and transform the MRE for $U$
[eq. \rf{r23.1}] into an equivalent MRE for $B$. As result, we
arrive at the following pair of MREs
\begin{eqnarray}
B^{\prime}&=&R^{-1}M_0 R-K_1^{-1}BB+BK_1^{\prime}
+\left[NN^{\#}+(N^{\#})^{\prime}\right]K_1,\label{n1.1}\\
B^{\prime}&=&-M_1+BK_1^{-1}B\label{n1.2}\;
\end{eqnarray}
which should be satisfied simultaneously.
 The corresponding consistency test will be performed in two
steps:
\begin{itemize}
\item[1.] From the limiting behavior at $r\to 0$ we will derive a
relation between $l_0$ and $l_1$.
\item[2.] We will extract from eqs. \rf{n1.1}, \rf{n1.2} a system of non-linear ODEs for
the helical turbulence functions $\alpha_0, \alpha_1$ and for the
 components $b_1,\ldots,b_4$ of the matrix $B$.
By mutual substitutions of these ODEs we will find an
inconsistency which can be interpreted as a no-go theorem.
\end{itemize}

\noindent {\bf \ref{no-go}.1 Limiting behavior at $r\to 0$}\\
{}From the assumed non-singular behavior of the helical turbulence
functions at $r\to 0$ follows that they can be approximated as
$$
\alpha_{0,1}(r\to 0)\approx c_{0,1}+a_{0,1}r+{\cal O}(r^2) \ ,
\qquad c_{0,1}\neq 0.
$$
Substituting this approximation in a slightly
rewritten version of the defining equation \rf{r27.2} for the
matrix  $Y$
\begin{equation}\label{n3}
\left[I\partial_r^2-\alpha_1^{\prime}\sigma_-\partial_r -
\frac{l_1(l_1+1)}{r^2}I-\left(\begin{array}{cc}E&-\alpha_1\\
\alpha_1&E-\alpha_1^2\end{array}\right)\right]Y=0
\end{equation}
we obtain the estimate
$$
Y(r\to 0)\approx r^{-l_1}\left(I+\frac{a_1}{2}\sigma_- r +
 {\cal O}(r^2)
\right)\left(r^{2l_1 +1}C_+ + C_- \right),
$$
where $C_+$, $C_-$ are arbitrary non-singular constant matrices
$\det (C_{\pm})\neq 0$. Correspondingly, it holds
\begin{eqnarray}
Z:=Y^{\prime}Y^{-1}&\approx &-l_1 r^{-1}I+\frac{a_1}{2}\sigma_-
+{\cal O}(r) \ , \label{n5.1}
\\ B=-K_1 Y^{\prime}Y^{-1}&\approx &l_1 r^{-1}I-\left[c_1l_1 r^{-1}+(l_1+1/2)a_1
\right]\sigma_- + {\cal O}(r) \ .\label{n5.2}
\end{eqnarray}
Comparison of \rf{n5.2} with \rf{r15} shows that the components
$b_2$ and $b_4$ of the matrix $B$ vanish at least as
$$
b_2,b_4\approx {\cal O}(r) \quad \textnormal{for} \ r\to 0 \ .
$$
Furthermore, we find with the help of eqs. \rf{r19.2}, \rf{r19.3} and \rf{r21}
that $q\approx {\cal O}(r)$ and, hence, $a_0/c_0=a_1/c_1$, as well as $q^{\prime},f,f^{\prime}\approx {\cal O}(1)$
what implies $N,N^{\#},(N^{\#})^{\prime}\approx {\cal O}(1)$.

We are now well prepared to perform a partial consistency test of
\rf{n1.1} and \rf{n1.2} by comparing the singular terms of these
equations in the vicinity of the origin $r=0$. From the MREs
\rf{n1.1} and \rf{n1.2} we find
\begin{eqnarray}
-K_1^{-1}K_1^{\prime}Z-Z^{\prime}&=&\frac{l_0(l_0+1)}{r^2} I-K_1^{-1}ZK_1Z-ZK_1^{\prime}
+{\cal O}(1)\label{n7.1},\\
-K_1^{-1}K_1^{\prime}Z-Z^{\prime}&=&-\frac{l_1(l_1+1)}{r^2}I+ZZ+{\cal
O}(1)\label{n7.2},
\end{eqnarray}
respectively. Substituting $Z$ from \rf{n5.1} and equating the coefficients of the
$r^{-2},r^{-1}-$terms we obtain from eq. \rf{n7.1}
$$
l_1=l_0+1, \qquad a_1=0
$$
and, hence, also $a_0=0$.
Eq. \rf{n7.2} is automatically satisfied, because $Y$ is defined
by the corresponding linearized equation \rf{n3}.
The incremental relation $l_1=l_0+1$ is well known from ladder operator constructions for
 spherically symmetric Hamiltonians in QM
\cite{raab}. This is not surprising, because
this ladder operator construction can be recovered from the
intertwining construction \rf{r1} for the $\alpha^2-$dynamo operator
matrices by the two-step transition: \
 1. $\alpha_0=\alpha_1=\alpha$,  2. $\alpha\to 0$.

\noindent {\bf \ref{no-go}.2 Systems of coupled non-linear ODEs
and their inconsistency}\\ The system of eight coupled non-linear
ODEs for the components $b_1,\ldots,b_4$ of the matrix $B$ is
easily obtained from the MREs \rf{n1.1}, \rf{n1.2}, e.g. with the
help of the matrix multiplication package of {\small
MATHEMATICA}$^\copyright$. For our analysis it is sufficient to
consider only the simplest four equations of this system, i.e. the
$\sigma_+$ and $I$ projections of \rf{n1.1} and \rf{n1.2}:
\begin{eqnarray}
b_2^{\prime}&=&2b_1b_2+\alpha_1(1+b_2^2)\label{n9.1}\\
&=&-2b_1b_2-\frac{\alpha_0^2}{\alpha_1}\label{n9.2}\\ \nn \\
b_1^{\prime}&=&b_1^2+b_2b_3-b_4^2-E-\frac{l_1(l_1+1)}{r^2}+\alpha_1b_1b_2\label{n9.3}\\
&=&-b_1^2-b_2b_3+b_4^2+E+\frac{l_0(l_0+1)}{r^2}-\alpha_1^{\prime}b_2+\frac{\alpha_0^2}{2}+q^{\prime}-q^2
.\label{n9.4}
\end{eqnarray}
Equating the right-hand-sides of \rf{n9.1}, \rf{n9.2} and using
$b_2=2q/\alpha_1$ from \rf{r21} we are able to express $b_1$ as
\begin{equation}\label{n10}
b_1=-\frac{4q^2+\alpha_0^2+\alpha_1^2}{8q}.
\end{equation}
Taking into account that $q=\partial_r \ln(\alpha_0/\alpha_1)/2$ according to \rf{r19.2} and that
the helical turbulence functions
$\alpha_0$ and $\alpha_1$ do not depend on $l_0$ or $l_1$ we
conclude from equation  \rf{n10} that $b_1$ should not depend on $l_0$ or
$l_1$ too.
On the other hand, addition of \rf{n9.3} and \rf{n9.4} together with the relation $l_0=l_1-1$
gives
$$
2b_1^{\prime}=-\frac{2l_1}{r^2}
+2q\left(b_1-\frac{\alpha_1^{\prime}}{\alpha_1}\right)+\frac{\alpha_0^2}{2}+q^{\prime}-q^2
$$
what by integration leads to a function $b_1$ which depends on
$l_1$. I.e. the term depending on $l_1$ cannot be compensated by a
combination of $l_1-$independent terms. This is an obvious
contradiction to \rf{n10} and we have to conclude that the
consistency conditions \rf{r6.1} - \rf{r6.6} cannot be fulfilled
simultaneously. This means that we are lead to the \\ {\bf No-go
theorem:}\\ The structure of the MHD $\alpha^2-$dynamo operator
matrix is incompatible with an operator intertwining technique
which is based on first-order differential intertwining operators.

A similar situation occurs also for three-dimensional spherically
symmetric models in QM \cite{raab}. There the $l-$dependent
centrifugal term sets so strong restrictions on the form of the
allowed potential that an intertwining construction built on
first-order differential intertwining operators is only possible
for the following three cases: the constant potential
$V(r)=\const$, the Coulomb potential $V(r)\propto 1/r$, and the
potential of the three-dimensional isotropic harmonic oscillator
with $V(r)\propto r^2$. Richer classes of allowed potentials are
only found for models in their $s$ states, when $l=0$. Such states
are {\it a priori} excluded  for the $\alpha^2-$dynamo operator
matrix due to its construction [see Eq. \rf{a7}].

\section{Concluding remarks\label{conclu}}

In the present paper, we have tested the MHD $\alpha^2-$dynamo
operator matrix for its compatibility with the simplest variant of
an intertwining construction based on {\it first-order}
differential intertwining operators. The operators have been
chosen in accordance with the fundamental $J-$symmetry
(pseudo-Hermiticity) of the operator matrix and lead to a set of
six matrix equations as consistency conditions. With the help of a
step-by-step reduction of the complexity we have extracted their
basic structural elements and have shown that they contain an
intrinsic inconsistency. So, we have to conclude that the
structure of the $\alpha^2-$dynamo operator matrix is not
compatible with the considered first-order differential
intertwining Ansatz. This fact is the subject of the formulated
no-go theorem.

It remains to test whether intertwining constructions can be built
from second-order or higher-order differential intertwining
operators.
Energy shift operators based on
second-order differential expressions are known for harmonic
oscillators with time-dependent frequencies and additional
$1/r^2-$term \cite{quadratic} as well as for the spherically
symmetric oscillator and the Coulomb potential \cite{raab}. A
generalization of the technique to the MHD $\alpha^2-$dynamo
operator matrix seems realistic.

Another approach for a clarification of the considered isospectrality
problem could consist in a generalization of the Gelfand-Levitan
technique for vector-valued Sturm-Liouville problems \cite{gelfand1}.
 Concerning its general structure, the $\alpha^2-$dynamo operator matrix $\hat
H_l[\alpha]$ is a singular non-self-adjoint matrix Sturm-Liouville
operator which by a unitary transformation can be recast into the
standard form
$$
-\partial_r P_2(r)\partial_r+ P_0(r)\; .
$$
In 1998, Jodeit and Levitan \cite{gelfand1} analyzed the isospectrality problem
for matrix Sturm-Liouville operators with  $P_2(r)=I$ and $P_0(r)$ a symmetric matrix.
They showed that if two vector-valued Sturm-Liouville problems are
isospectral then the eigenfunctions of one problem can be constructed from the eigenfunctions of the other problem
with the help of a matrix Gelfand-Levitan transformation.
   So, a generalization of this technique to Sturm-Liouville problems with
 non-symmetric $P_0(r)$ and $P_2(r)\neq
I_2$ would naturally cover the isospectrality problem for the MHD
$\alpha^2-$dynamo operator matrix.\\[2ex]
\noindent {\bf Acknowledgements}\\
We would like to thank G. Gerbeth for numerous discussions and
C. Tretter for useful comments.
This project was supported by
the German Research Foundation, DFG, under grant GE-682/12-1.
\appendix
\section{Derivation of the $\alpha^2-$dynamo operator matrix from
the mean-field induction equation\label{induct1}}
\renewcommand{\theequation}{\thesection\arabic{equation}}
\setcounter{equation}{0}

For completeness we sketch here the main steps of the derivation
of the $2\times 2$ operator matrix $ \hat H_l[\alpha] $ for a
model with helical turbulence function $\alpha(r)$. The outline
follows the technique for models with $\alpha
=\const $ as presented in \cite{krause1}.

The spherical MHD mean-field $\alpha^2-$dynamo in its kinematic
regime is described by the induction equation for the magnetic
field
\begin{equation}\label{a1}
\partial_t \vect{B}=\bsym{\nabla \times }(\alpha \vect{B})
+\nu_m \Delta \vect{B}
\end{equation}
supplemented by the condition $\bsym{\nabla }\bsym{ \cdot} \vect{B}=0$.
The magnetic diffusivity $\nu_m$ is assumed to be constant and the
helical turbulence function $\alpha$ to depend only on the
distance from the origin $\alpha=\alpha(r)$. Decomposition into
toroidal and poloidal components $\vect{B}=\vect{B_t}+\vect{B_p}$
and setting $\vect{B_p}=\bsym{\nabla \times }\vect{A_t}$
allows for a decomposition of the induction equation \rf{a1}
\begin{eqnarray}
\partial_t \vect{B_t}&=&\bsym{\nabla \times }(\alpha \bsym{\nabla \times }\vect{A_t})
-\nu_m \bsym{\nabla \times }\bsym{\nabla \times } \vect{B_t}\label{a2.1}\\
\partial_t \vect{A_t}&=&\alpha \vect{B_t}
-\nu_m \bsym{\nabla \times }\bsym{\nabla \times }
\vect{A_t}.\label{a2.2}
\end{eqnarray}
Furthermore, the fields $\vect{B_t}$ and $\vect{A_t}$ can be
represented as
$$
\vect{A_t}=-\vect{r}\bsym{\times \nabla} F_1, \qquad
\vect{B_t}=-\vect{r}\bsym{\times } \bsym{\nabla } F_2,
$$
where $F_1$ and $F_2$ are single-valued scalar functions which are
normalized on the unit sphere $S^2$ by the condition
\begin{equation}\label{a3a}
\int_{S^2}F_{1,2} d\omega =0.
\end{equation}
With the help of the relations
\begin{eqnarray*}
\Delta (\vect{r}\bsym{\times \nabla}F_1)&=&\vect{r}\bsym{\times
\nabla}\Delta F_1 \nn \\ \bsym{\nabla \times }\left[\alpha
\bsym{\nabla \times }( -\vect{r}\bsym{\times } \bsym{\nabla
}F_1)\right]&=&\vect{r}\bsym{\times \nabla}\left[\frac 1r
(\partial_r \alpha)(\partial_r rF_1)+\alpha \Delta F_1\right],
\nn\\ \alpha \vect{r}\bsym{\times } \bsym{\nabla } F_2&=&
\vect{r}\bsym{\times } \bsym{\nabla } (\alpha F_2)\nn
\end{eqnarray*}
equations \rf{a2.1} and \rf{a2.2} can be rewritten
as
\begin{eqnarray*}
\vect{r}\bsym{\times \nabla}\left[\nu_m \Delta F_1+\alpha
F_2-\partial_t F_1\right]&=&0, \nn\\ \vect{r}\bsym{\times
\nabla}\left[\nu_m \Delta F_2-\frac 1r (\partial_r
\alpha)(\partial_r r F_1) -\alpha \Delta F_1-\partial_t
F_2\right]&=&0. \nn
\end{eqnarray*}
It follows that the
expressions in the square brackets are functions of $r$ and $t$
alone which must vanish due to the normalization condition
\rf{a3a} and its implication $\int_{S^2}\Delta F_{1,2} d\omega
=0$. By re-scaling of $r$ and $t$ one sets the magnetic
diffusivity
to unity $\nu_m=1$ and the boundary conditions at $r=1$.

With the help of a series expansion in spherical harmonics
$$
F_{1,2}=\sum_{l,m,n}e^{\lambda_{l,n} t} F_{1,2}^{(l,m,n)}(r)Y^m_l
(\theta, \phi) \in L^2(\Omega, r^2dr)\otimes L^2(S^2, d\omega),
\quad \Omega =[0,1]
$$
one obtains the eigenvalue problem
\begin{eqnarray*}
\Delta_l F_1^{(l,m,n)}+\alpha F_2^{(l,m,n)}&=&\lambda_{l,n}
F_1^{(l,m,n)} \nn\\ \Delta_l F_2^{(l,m,n)} -\frac 1r (\partial_r
\alpha)(\partial_r r F_1^{(l,m,n)}) -\alpha \Delta_l
F_1^{(l,m,n)}&=&\lambda_{l,n} F_2^{(l,m,n)}. \nn
\end{eqnarray*}
Here we used the notation $\Delta_l =\frac{1}{r^2}\partial_r
r^2\partial_r -\frac{l(l+1)}{r^2}$ and the fact that due to the
symmetry of the dynamo configuration \cite{krause1} the eigenvalues
$\lambda_{l,n}$ depend only on $l$ and $n$. We note that the
normalization condition \rf{a3a} implies
\begin{equation}\label{a7}
F_{1,2}^{(l=0,m,n)}=0.
\end{equation}
Finally, the substitutions $p=-i(\partial_r+1/r)$, \
$\psi_{1,2}=F_{1,2}^{(l,m,n)}\in L^2(\Omega, r^2dr)$ lead to the
eigenvalue problem for the $\alpha^2-$dynamo operator matrix $\hat
H_l[\alpha]$ as it is given in eq. \rf{i2} of the Introduction.

%
\end{document}